\documentclass[twocolumn, aastex]{emulateapj}
\usepackage{amssymb}
\usepackage{amsmath}
\usepackage{mathrsfs}
\usepackage{amsfonts}
\usepackage{textcomp}
\usepackage{color}
\definecolor{mycol}{cmyk}{0.0, 0.11, 0.12, 0.0}
\definecolor{mycol}{cmyk}{0.11, 0.99, 0.11, 0.0}

\def\apjl{\mbox{ApJL}}
\def\apjs{\mbox{ApJS}}

\def\aap{\mbox{A\&A}}

\def\deg{^\circ}

\def\cm{cm$^{-3}$}
\def\kms{km~s$^{-1}$}

\def\deg{^\circ}

\def\farcs{\hbox{$.\!\!^{\prime\prime}$}}
\def\Ha{H$\alpha$}

\def\Sii{\hbox{[{\sc Sii}]}}
\newcommand{\Msun}{M$_\odot$}
\newcommand{\Msunyr}{M$_\odot$\,yr$^{-1}$}
\newcommand{\Mdot}{\dot{M}}

\shorttitle{ALMA follows streaming of dense gas down to 40 pc from the supermassive black hole in NGC\,1097}
\shortauthors{Fathi et al.}

\slugcomment{Accepted for Publication in the Astrophysical Journals Letters on April 24, 2013}
\begin{document}

\title{ALMA follows streaming of dense gas down to 40 pc from the supermassive black hole in NGC\,1097}

\author{Kambiz Fathi$^{1,2}$}
\author{Andreas A. Lundgren$^{3}$} 
\author{Kotaro Kohno$^{4,5}$} 
\author{Nuria Pi\~nol-Ferrer$^{1}$}
\author{Sergio Mart\'{\i}n$^{6}$}
\author{Daniel Espada$^{7}$} 
\author{Evanthia Hatziminaoglou$^{8}$}
\author{Masatoshi Imanishi$^{9}$}
\author{Takuma Izumi$^{4}$} 
\author{Melanie Krips$^{10}$} 
\author{Satoki Matsushita$^{11}$}
\author{David S. Meier$^{12,13}$}
\author{Naomasa Nakai$^{14}$}
\author{Kartik Sheth$^{15}$} 
\author{Jean Turner$^{16}$}
\author{Glenn van de Ven$^{17}$}
\author{Tommy Wiklind$^{3}$}
\affil{\ \\ $^{1}$Stockholm Observatory, Department of Astronomy, Stockholm University, AlbaNova Centre, 106 91 Stockholm, Sweden}
\affil{$^{2}$Oskar Klein Centre for Cosmoparticle Physics, Stockholm University, 106 91 Stockholm, Sweden}
\affil{$^{3}$Joint ALMA Observatory, Alonso de C\'ordova 3107, Vitacura - Santiago, Chile}
\affil{$^{4}$Institute of Astronomy, The University of Tokyo, 2-21-1 Osawa, Mitaka, Tokyo 181-0015, Japan}
\affil{$^{5}$Research Center for the Early Universe, The University of Tokyo, 7-3-1 Hongo, Bunkyo, Tokyo 113-0033, Japan}
\affil{$^{6}$ESO, Alonso de C\'ordova 3107, Vitacura - Santiago, Chile}
\affil{$^{7}$National Astronomical Observatory of Japan (NAOJ), 2-21-1 Osawa, Mitaka, Tokyo 181-8588, Japan}
\affil{$^{8}$ESO, Karl-Schwarzschild-Str. 2, 85748 Garching bei M\"unchen, Germany}
\affil{$^{9}$Subaru Telescope, National Astronomical Observatory of Japan, 650 North A'ohoku Place, Hilo, Hawaii 96720, USA}
\affil{$^{10}$Institute for Radio-Astronomy at Millimeter Wavelengths, Domaine Univ., 300 Rue de la Piscine, 38406 Saint Martin d'Heres, France}
\affil{$^{11}$Academia Sinica, Institute of Astronomy and Astrophysics,
 P.O. Box 23-141, Taipei 10617, Taiwan, R.O.C.}
\affil{$^{12}$Department of Physics, New Mexico Institute of Mining and Technology, 801 Leroy Place, Socorro, NM 87801, USA}
\affil{$^{13}$National Radio Astronomy Observatory, P.O. Box O, Socorro, NM 87801, USA}
\affil{$^{14}$Division of Physics, Faculty of Pure and Applied Science, University of Tsukuba, Tsukuba, Ibaraki 305-8571, Japan}
\affil{$^{15}$National Radio Astronomy Observatory, 520 Edgemont Road, Charlottesville, VA 22903, USA}
\affil{$^{16}$Department of Physics and Astronomy, UCLA, Los Angeles, CA 90095-1547, USA}
\affil{$^{17}$Max Planck Institute for Astronomy, K\"onigstuhl 17, 69117 Heidelberg, Germany}

\begin{abstract}
We present a kinematic analysis of the dense molecular gas in the central 200 parsecs of the nearby galaxy NGC\,1097, based on Cycle 0 observations with the Atacama Large Millimeter/sub-millimeter Array (ALMA). We use the HCN(4--3) line to trace the densest interstellar molecular gas ($n_{\rm{H}_2} \sim 10^8$ \cm), and quantify its kinematics, and estimate an inflow rate for the molecular gas. We find a striking similarity between the ALMA kinematic data and the analytic spiral inflow model that we have previously constructed based on ionized gas velocity fields on larger scales. We are able to follow dense gas streaming down to 40 pc distance from the supermassive black hole in this Seyfert~1 galaxy. In order to fulfill marginal stability, we deduce that the dense gas is confined to a very thin disc, and we derive a dense gas inflow rate of $0.09$ \Msunyr\ at 40 pc radius. Combined with previous values from the \Ha\ and CO gas, we calculate a combined molecular and ionized gas inflow rate of $\sim 0.2$ \Msunyr\ at 40 pc distance from the central supermassive black hole of NGC\,1097.
\end{abstract}

\keywords{
galaxies: kinematics and dynamics ---
galaxies: active ---
galaxies: individual (NGC\,1097)}

\section{Introduction}\label{intro}
The central region of the nearby barred Seyfert 1 galaxy NGC\,1097 displays a number of intriguing morphological and kinematic features. At $\sim 1 $ kpc, an almost circular ring-like feature marks the transition between the prominent $R\sim 8$ kpc galactic bar and the relatively diffuse region interior to the ring. The bar hosts two prominent dust lanes, both originating at around the Corotation radius of the bar \citep{npf13}, cutting through the inner ring and transforming into nuclear spirals that continue down to $\sim3.5$ pc distance from the active nucleus \citep{lou01,f06}. The dust lanes are accompanied by diffuse ionized gas revealing clear kinematic signatures of bar-induced gas inflow over the entire face of the galaxy \citep{npf13}. 

Neutral gas and ionized gas data cubes \citep{o89,f06} have confirmed the `abundant' presence of these two phases of the interstellar medium across the central kpc radius. However, interferometric molecular gas maps show emission confined to the nuclear ring and the central 2-3 hundred parsecs radius \citep{k03,h08,h11,h12}. Moreover, \cite{npf11} showed that the bulk of the interstellar gas at the centre of NGC\,1097 (like in many other galaxies) is in the molecular phase, and therefore, a detailed analysis of the central gas concentration needs to account for the different physical conditions. The interplay between the different phases provide crucial clues to understanding the energies involved in redistributing the gas in a way that leads to the observed phase transition efficiencies.

The discovery of broad ($\sim 10000$ km/s) double-peaked H$\alpha$ emission lines by \citet{sb93} makes NGC\,1097 also an ideal laboratory for studying the fate of the gas accumulated in the centers of active galactic nuclei (AGN). At a distance of 14.5 Mpc (i.e., $\sim 70$ pc/\arcsec), this galaxy is also suitable for high-resolution studies of the physical processes that cause the material/fuel to loose its angular momentum and fall toward the AGN \citep[e.g.,][]{sb03}. 

Although it is straightforward to transport gas down to the central kpc and induce enhanced star formation, it is more difficult to make the gas reach smaller scales (few pc) required to fuel an AGN. In rotating systems, perturbations can cause the potential to become non-axisymmetric, and torques exerted by the subsequent non-axisymmetric features are able to drive material toward the centre \citep{schwarz84}. \cite{st99} have argued that magnetic stress may aid the infalling gas to complete the last few parsecs down to the central supermassive black hole (SMBH). 

To build a realistic scenario for the fate of the gas that is piling up around an AGN to eventually fuel it, one has to make a detailed analysis of the distribution and kinematics of multiple phases of the interstellar gas in the region of interest.

Based on ionized gas kinematic maps (van de Ven \& Fathi 2010, hereafter {\bf vdVF}, and Pi\~nol-Ferrer et al. 2013), we have derived a concise picture for NGC\,1097, according to which, the gravitational perturbation that once gave rise to the formation of the prominent bar drives the evolution of structure inside the bar as well as the outer spiral arms. The outer spiral arms are confined between the Corotation radius and reach beyond the Outer Lindblad Resonance radius of the main galactic bar. The circumnuclear ring once formed at the location of the Outer Inner Lindblad resonance \citep{npf13}, and has likely migrated toward the centre of the galactic gravitational potential \citep{rt03,vdv09}. Inside the ring, the non-circular velocities are consistent with the presence of two spiral arms (in morphology). The nuclear arms are well disguised in optical images, and several image-enhancement techniques have led different authors to argue for a different number of arms \citep[e.g.,][and  vdVF]{lou01,d09}. Most of the disagreements concern the inner $\sim 100$ pc, where increased dust content may have distorted both images and kinematic measurements in the optical and near infrared. 

Here we present a quantitative analysis of the kinematics of the densest molecular gas within the central kpc radius of NGC\,1097 based on Atacama Large Millimeter/sub-millimeter Array observations of Hydrogen cyanide, HCN(4--3) (Proposal 2011.0.00108.S, PI: Kohno). We compare the ALMA kinematic data with a dynamical model that we have previously constructed based on two-dimensional spectroscopic data of ionized gas.

\begin{figure*}
\centering\includegraphics[width=0.98\textwidth]{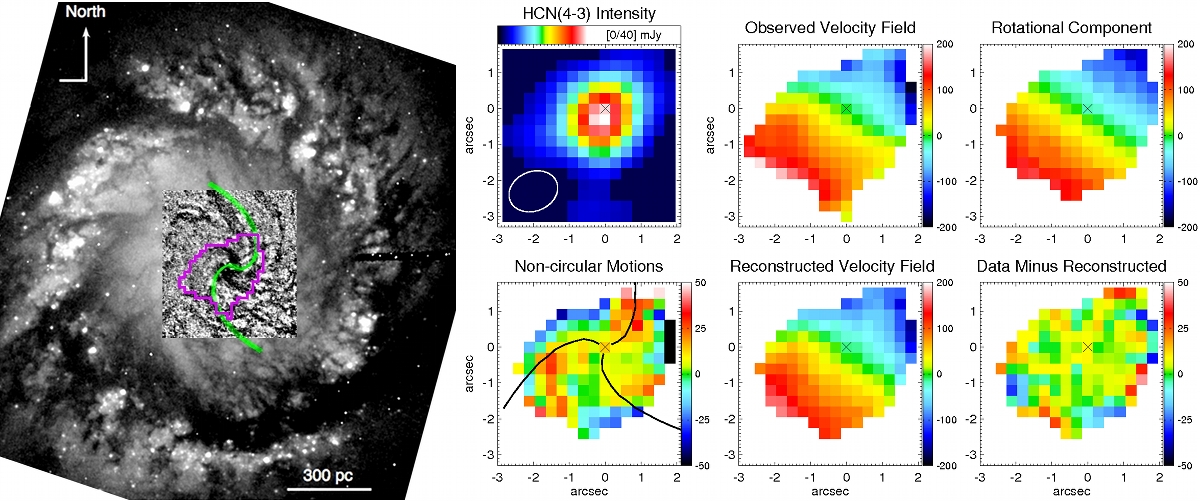}
\caption{{\em Left:} Hubble Space Telescope H$\alpha$ image with a structure map showing the nuclear dust features \citep{f06} and the footprint of the ALMA HCN(4--3) pixels with $A/N\ge 20$ (magenta contours). Overplotted, in green, are the spiral arms modeled by vdVF. {\em Right:} ALMA HCN(4--3) distribution and velocity field. The circular and non-circular motion maps, Fourier reconstructed and final residual maps are displayed in \kms. The black curves in the bottom left panel outline the predicted location of the non-circular motions due to the model spiral arms corresponding to the two-arm morphological arms drawn on the {\em HST} image (in green). The non-circular velocities follow a similar three-fold symmetry. The ellipse at the top left panel displays the  beam size, and the cross marks the kinematic centre, offset from the peak centroid by less than 0\farcs3. }
\label{fig:1}
\end{figure*}

\section{The data}\label{sec:data}
NGC\,1097 was observed with ALMA on November 5 and 6, 2011, with 14 and 15 antennas, respectively. Our Band 7 observations targeted the HCN(4--3) line at rest frequency of 354.505 GHz with the original channel spacing of 488.28125 kHz. To increase the signal strength, the data were binned by a factor 20, leading to an effective channel width of 9.77 MHz ($\sim 8.3$ \kms). The primary beam was $18\farcs1$ with a synthesized beam of $1\farcs50\times 1\farcs20$ ( $\sim 105 \times 84$ pc) and at $-72.4\deg$ position angle, sampled at $0\farcs3$/pix. All data specifications are described in \cite{i13}. 

Figure~\ref{fig:1} shows the integrated intensity map of HCN(4--3) with the velocity moment maps and models, described below. At this resolution the spiral arms are difficult to see in the integrated intensity map, which sums all velocities, but are easier to detect in the kinematics. 

The HCN(4--3) kinematics were derived in two ways. Simple first Moment map (intensity-weighted mean velocity map) was cross checked with Gaussian fitting to each individual spectrum. We found no significant signatures of non-Gaussianity, however, the Gaussian fits resulted in a generally noisier velocity field. We use the Moment 1 maps and exclude all pixels for which the corresponding spectrum Amplitude-over-Noise $A/N<20$ (see Fig.~\ref{fig:1}). We find almost no signal outside the area illustrated in Fig.~\ref{fig:1} at lower $A/N$ values. This confirms that the dense gas is confined to a small region around the AGN in NGC\,1097 \citep{h12}.

The ratio between the Einstein coefficient and collisional rates for the HCN(4--3) transition results in an estimated critical density of a few times $10^7$ to $10^8$ \cm\ with a small dependency on temperature. Our calculation is conformal with previously reported densities of $n_{\rm{H}_2}\sim 10^8$ \cm\ at 40 K kinetic temperature \citep[e.g.,][]{choi00, t07}. Such high densities are also consistent with the deep obscuration expected toward the center of a galaxy \citep[e.g.,][]{k96,s10}.

To constrain the large scale gas kinematics, we have used a mosaic of Fabry-Perot interferometric observations at 0\farcs83 spatial sampling, covering a $7\arcmin\times 7\arcmin$ field at 15~\kms\ spectral resolution \citep{d08,npf13}. To further look into the central few 100 parsecs, we used \Ha\ two-dimensional velocity field, at 0\farcs1 spatial sampling and 85~\kms\ spectral resolution, from the Gemini South Telescope's Integral Field Unit, covering the inner $7\arcsec\times 15\arcsec$ \citep{f06}. The combination of the two sets of data is imperative, as they present a coherent dynamical model for NGC\,1097 from $\sim 20$ kpc down to $\sim 100$ pc from its central SMBH.

\section{Kinematic Parameters}\label{sec:analysis}
We assume that the galactic disc is predominantly rotating and apply the prescription used in vdVF to quantify the velocity field shown in the top middle panel of Fig.~\ref{fig:1}. We divide the observed velocity field into concentric rings, each containing $>7$ pixels (this sets the limit for the innermost radius to $0\farcs55 \lesssim 40$ pc). We fix the inclination at $35\deg$ and apply a $\chi^2$ minimization to obtain the central coordinates $(x_0, y_0)$, systemic velocity $V_{\rm sys}$ and position angle PA of the disc. After fixing these parameters, we make the final $\chi^2$ fit to the desired modes (here up to and including 3rd order) of the Fourier decomposition mathematically formulated as
\begin{equation}
V_{\rm los} = V_{\rm sys} + \sum_{n=1}^k [c_n(r) \cos n\theta + s_n(r) \sin n\theta] \sin i .
\label{eq:vlos}
\end{equation}

We apply Eq.~(\ref{eq:vlos}) to the HCN(4--3) velocity field and find an agreement between the kinematic and photometric center to within one pixel (see Fig.~\ref{fig:1}). Each ring is populated at more than 225 degrees, hence, we ensure that we do not need to assume any level of symmetry in the observed non-circular motions. We derive a systemic velocity of $1300\pm48$ \kms\ and the kinematic position angle of $147\pm6\deg$. In Fig.~\ref{fig:1} we illustrate the different stages of the fits described here, and a three-fold symmetry can be found in the non-circular velocity field (as predicted for this region in NGC\,1097 by vdVF). The avreage final residual velocities are 10 \kms, indicating that we have reproduced 95\% of the observed velocity features with the Fourier decomposition method. 

\begin{figure}
\centering\includegraphics[width=0.49\textwidth]{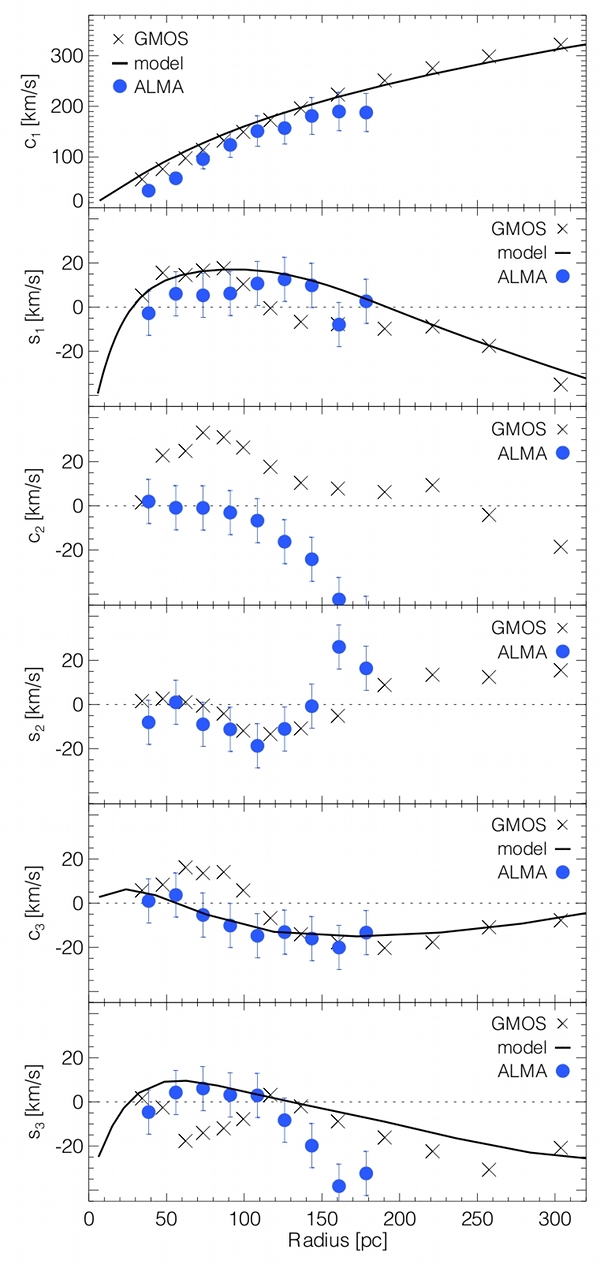}
\caption{Comparing the quantified kinematics of the ionized gas in the optical (crosses) and the HCN(4--3) ALMA data (filled circles) to the analytic model prediction presented in vdVF. The spiral model is not a fit to any of the two data sets shown here, but a carefully chosen set of spiral parameters that best match the GMOS data. The HCN(4--3) data points are simple overplots of the new ALMA data showing a striking agreement with the model predictions. The model does not contain $c_2$ or $s_2$ terms (see Eqs~(A12) and (A13) in vdVF).}
\label{fig:3}
\end{figure}

Fitting Gaussians to each spectrum gives us the average velocity uncertainty for all the pixels at 15 \kms. These errors are used to derive the uncertainties for the derived kinematics by means of Monte Carlo simulations. Repeated application to the Gaussian-randomized velocity field yields the uncertainties on the Fourier parameters. Our simulations show that the average rotation curve uncertainty is 20\% \citep[see also][]{f05}. Similarly, the higher Fourier term uncertainties have been calculated, and we plot the first, second and third Fourier terms of the HCN(4--3) data together with those derived from the GMOS data in Fig.~\ref{fig:3}.

\section{Analysis}
The ionized gas velocity Fourier terms were modeled by vdVF, who constrained the detailed structure of the nuclear spiral arms and the associated gas inflow from kinematic data (solid curves in the lower panels of Fig.~\ref{fig:3}). Their nuclear spiral structure is consistent with a weak perturbation in the gravitational potential due to a two-arm logarithmic spiral (in morphology) with a pitch angle of $52 \pm 4\deg$ derived directly from the Fourier expansion of the model velocity field. Similarly large pitch angles in the very central parts of galaxies have also been modeled by \cite{yy06}. Furthermore, the innermost $\sim 100$ pc radius of the data points analyzed by vdVF displayed the largest errors in the third Fourier terms, as a consequence of dust contamination. 

We test the effect of beam smearing on the derived kinematic parameters by artificially smoothing by a factor two, and find that the Fourier terms remain virtually unchanged. A notable effect of beam smearing is that it may lead to incorrect kinematic center, which in turn may cause uncertainties in the third Fourier terms. \cite{w00} found that smearing of 10\arcsec could produce up to 10 \kms\ third Fourier terms. This is less than our error bars. In light of these tests and following the discussion in \cite[][Chapter 2]{w00}, it unlikely that the innermost values would be affected by beam smearing. 

In Fig.~\ref{fig:3}, we overplot the ALMA HCN(4--3) velocity Fourier components on those derived from the GMOS data. The top panel reveals a very good agreement between the rotation curve ($\sim c_1$ in Eq~1) of the HCN(4--3) and that of the H$\alpha$ gas in the central 200 pc. Hence, due to the agreement of their overall kinematics, the bulk rotation of the dense gas is co-planar to the ionized interstellar gas. Further light can be shed on the dynamical behaviour of the HCN(4--3) by considering the higher Fourier terms. Figure.~\ref{fig:3} illustrates the analytic spiral model that has lead to the kinematic derivation of the pitch angle of the nuclear spirals with associated gas inflow rates \citep[vdVF and][]{npf11}. A simple overplot of the ALMA data (filled circles) displays the striking agreement between the relatively non-contaminated HCN(4--3) kinematic behaviour and the model predictions. {\em The analytic spiral model predicts the HCN(4--3) velocity Fourier terms down to $0\farcs55 \lesssim 40$ pc.}

Besides the $n=1$ and $n=3$ Fourier terms in the velocity field, the non-circular motions also contain marginal $n=2$ terms. vdVF argued that the $n=2$ terms in the H$\alpha$ were most likely due to dust contamination and possible shocks associated with the gas streaming along the nuclear spiral arms \cite[see also][]{v04}. This seems to be confirmed here with the HCN(4--3) second terms consistent with zero inside 100 pc radius.

\begin{figure}
\centering\includegraphics[width=0.49\textwidth]{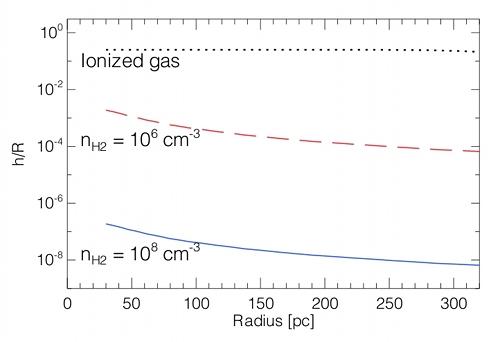}
\caption{Disc scale height over radius $h/R$ as a function of radius for the HCN(4--3) emitting gas at $n_{\rm{H}_2}\sim 10^8$ (solid curve). For comparison, we also illustrate this parameter for the ionized gas (dotted curve) and the molecular counterparts at lower densities (dashed curve).}
\label{fig:4}
\end{figure}

\section{Dense Gas Streaming Down to $\lesssim 40$ pc From the SMBH}
We use the derived $c_1$ curve to measure the dynamical mass, assuming a thin disc model and velocity uncertainties of $20\%$. We derive the mass within the central 100 pc to $3.5^{+1.6}_{-2.0}\times 10^8$ \Msun, and inside the 40 pc radius, $8.0^{+2.9}_{-3.5}\times10^6$ \Msun. These masses are based on velocity measurements well outside the sphere of influence of the SMBH \citep[$\sim 13$ pc,][]{p72}. Even so, in theory, a significant contribution from the SMBH is expected on the rotating velocities at 40 pc radius. However, the synthesized beam size of the ALMA data almost entirely smears out the effect of the SMBH at this radius, and therefore, a rotation curve at higher spatial resolution is needed to measure the dynamical mass of the SMBH. Furthermore, it should be noted that our mass estimate from the observed cold molecular gas rotation curve are not corrected for the contribution of non-cirular motions in the $c_1$ curve (see Fig.~3 in vdVF). Thus, these are likely lower mass limits.

The prescription for deriving mass inflow rates along the nuclear spirals has been presented in vdVF. They combined the resulting inflow velocity corresponding to the model spiral arm parameters, with the gas density in the spiral arms inferred from \Sii\ emission line ratios. They calculated the ionized gas inflow rate as a function of radius, reaching $0.033$ \Msunyr\ at a distance of 100 pc from the central SMBH. The inflow rate was later refined by \cite{npf11} who used CO gas measurements to derive one order of magnitude higher molecular gas inflow rate for a marginally stable disc model. Using this formalism, \cite{npf11} derived the CO gas inflow of $0.3$\Msunyr\ at 100 pc radius. 

\begin{figure}
\centering\includegraphics[width=0.49\textwidth]{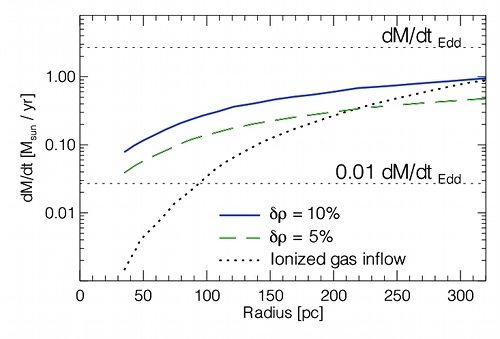}
\caption{Dense gas inflow rate derived from the analytic spiral model of vdVF as a function of galactocentric radius. Changing the molecular gas density by two orders of magnitudes does not change the inflow rate curves, however, the arm versus inter-arm overdensity $\delta \rho$ changes the inflow rate as illustrated by the sold and dashed curves.}
\label{fig:5}
\end{figure}

As we compare the HCN(4--3) kinematics with our previous results for ionized and CO gas \citep[vdVF,][]{npf11}, we set the sound speed at $10$ \kms\ and derive the dense gas scale height (Fig.~\ref{fig:4}) using the observed epicyclic frequency. We then use the kinematic parameters from the analytic spiral model to trace the gas inflow down to the resolution limit ($0\farcs55 \lesssim 40$ pc). The data set at hand does not allow investigating density difference between the nuclear spiral arms and the inter-arm region in NGC\,1097. We assume $10\%$ overdensity, similar to the upper limit of the observed value from the \Sii\ doublet, and in agreement with numerical simulations \citep{es00}. We derive the HCN(4--3) inflow rate of $0.3$ \Msunyr\ at 100 pc, and $0.09$ \Msunyr\ at 40 pc from the SMBH (see Fig.~\ref{fig:5}). The marginal stability criterion of \cite{r94} ensures that if the gas reaches higher densities, it will be confined to a thinner disc. Hence, changing the gas density by two orders of magnitude will not change the mass inflow rate (Fig.~\ref{fig:4}). However, accounting for the minimum density variation (arm versus inter-arm) derived from the \Sii\ line  ratios (i.e., $\delta \rho = 5\%$), the inflow rate would decreased by a factor 2 (Fig.~\ref{fig:5}).

The critical value $0.01~\dot{M}_{\rm Edd}$ for the transition between LINER and Sy1 galaxies was found by \citet{ho05} with over three orders of magnitudes distribution. The value that we have calculated here for the dense gas streaming, at 40 pc distance from the SMBH in NGC\,1097 corresponds to $\Mdot \sim 0.033$ $\dot{M}_{\rm Edd}$, where the Eddington accretion rate is onto a SMBH mass of $1.2 \times 10^8$ \Msun\ \citep[based on the central stellar velocity dispersion adopted from][and the $M_\bullet$--$\sigma$ relation of Tremaine et al. 2002]{l06}.

\section{Conclusions}
We have presented a detailed kinematics analysis of the dense interstellar gas in the circumnuclear region of the nearby Seyfert~1 galaxy NGC\,1097. We used ALMA Band~7 high-resolution observations of the HCN(4--3) emission line, which is postulated to trace $n_{\rm{H}_2}\sim 10^8$ \cm\ at 40 K kinetic temperature. While visual signatures of a rotation pattern dominate the observed velocity field, we have successfully applied Fourier decomposition of the velocity field to unveil the kinematic signatures of two prominent spiral arms (in morphology). The number of nuclear spiral arms in NGC\,1097 has been subject to a controversy and in the past years it has been clear that the presence of strong obscuration by dust at the very centre of this galaxy has complicated matters. We argue here that using velocity information from the HCN(4--3) clarifies matters in an unprecedented way.

We have found a striking agreement between the kinematics of the HCN(4--3) and an analytic spiral model that we previously build using ionized gas kinematic data at similar spatial resolution. The new ALMA data confirm that the spiral arms have pitch angle of $52\pm4\deg$, down to $\lesssim 40$ pc from the SMBH in NGC\,1097. 

We note that some studies have found that the HCN molecule could arise from radiation and vibrational excitation \citep{ld96,i04,mt12}. The relative contribution of the radiation pumping could then decrease the gas density with which the HCN is associated, and we note that an assumed density decrease by two orders of magnitudes will yield $h/R < 0.01$. For a thin disc, we have then measured the mass inside the 100 pc radius to be $3.5^{+1.6}_{-2.0}\times 10^8$ \Msun, and inside the 40 pc radius, $8.0^{+2.9}_{-3.5}\times10^6$ \Msun. The derived mass at 40 pc agrees with the $<3.0\times10^7$ \Msun, derived from the observed line flux following the procedure described in \citet{gs04}.

We have used a constant arm versus inter-arm overdensity of 10\% and kinematically derived a dense gas inflow of $0.3$ \Msunyr\ at 100 pc and $0.09$ \Msunyr\ at 40 pc radius. In combination with our previously derived values from the ionized and CO gas, we calculate a  molecular and ionized gas infall of $0.6$ \Msunyr\ at 100 pc and $\sim 0.2$ \Msunyr\ at 40 pc distance from the central SMBH of NGC\,1097. This inflow corresponds to $\Mdot \sim 0.066 \dot{M}_{\rm Edd}$ onto a black hole in NGC\,1097 with a mass of $1.2 \times 10^8$ \Msun. From the current data, it is not clear how much, if any at all, of the observed HCN(4--3) is radiationally excited by the active nucleus. In the presence of radiational excitation, the dense gas scale height that we present here will be lower limits. Notwithstanding, the gas inflow rates remain unchanged.

\section*{ACKNOWLEDGMENTS}
We thank the referee whose comments improved our manuscript. 
This paper makes use of the following ALMA data: 
ADS/JAO.ALMA\#2011.0.00108.S (PI: Kohno). ALMA is a partnership of ESO, NSF (USA) and NINS (Japan), together with NRC (Canada) and NSC and ASIAA (Taiwan), in 
cooperation with the Republic of Chile. The Joint ALMA Observatory is 
operated by ESO, AUI/NRAO and NAOJ. The NRAO is a facility of the National Science Foundation operated under cooperative agreement by Associated Universities, Inc.. Sergio Mart\'{\i}n is cofunded under Marie Curie Actions of the EC (FP7-COFUND). Satoki Matsushita is supported by NSC 100-2112-M-001-006-MY3 of Taiwan. Kambiz Fathi acknowledges support from the Swedish Research Council \& the Swedish Royal Academy of Sciences' Crafoord Prize Foundation.

\label{lastpage}
\end{document}